\documentclass[%
superscriptaddress,
preprint,
mathtools,
aps, %prl,
pra,
]{revtex4-1}

\usepackage{graphicx}% Include figure files
\graphicspath{{Figures/}}
\usepackage{epstopdf}
\usepackage{dcolumn}% Align table columns on decimal point
\usepackage{bm}% bold math
\usepackage{comment}
\usepackage{color}
\usepackage{physics}
\usepackage{siunitx}
\usepackage{chemformula}
\usepackage{ragged2e}
\usepackage[flushleft]{threeparttable}
\usepackage{hyperref} % add hypertext capabilities
\hypersetup{
    colorlinks  = true,
    citecolor   = blue,
    linkcolor   = blue,
    urlcolor   =blue
}
\begin{document}

\title{Multi-domain Characterization of Ferroelectric Switching Dynamics with a Physics-based SPICE Circuit Model for Phase Field Simulations}% Force line breaks with \\

\author{Chia-Sheng Hsu}
\affiliation{School of Electrical and Computer Engineering, Georgia Institute of Technology, Atlanta, GA 30332, USA
}%
\email{chiasheng@gatech.edu}
% \altaffiliation[Also at ]{Department of Electrical and Computer Engineering, Georgia Institute of Technology}
% Lines break automatically or can be forced with \\
\author{Sou-Chi Chang}%
\affiliation{Components Research, Intel Corporation, Hillsboro, OR 97124, USA}%
\author{Dmitri E. Nikonov}%
\affiliation{Components Research, Intel Corporation, Hillsboro, OR 97124, USA}%
\author{Ian A. Young}%
\affiliation{Components Research, Intel Corporation, Hillsboro, OR 97124, USA}%
\author{Azad Naeemi}%
\affiliation{School of Electrical and Computer Engineering, Georgia Institute of Technology, Atlanta, GA 30332, USA
}%

\date{\today}% It is always \today, today, but any date may be explicitly specified.

\begin{abstract}
In this paper, the multi-domain nature of ferroelectric (FE) polarization switching dynamics in a metal-ferroelectric-metal (MFM) capacitor is explored through a physics-based phase field approach, where the three-dimensional time-dependent Ginzburg-Landau (TDGL) equation and Poisson's equation are self-consistently solved with the SPICE simulator. Systematically calibrated based on the experimental measurements, the model well captures transient negative capacitance in pulse switching dynamics, with domain interaction and viscosity being the key parameters. It is found that the influence of pulse amplitudes on voltage transient behaviors can be attributed to the fact that the FE free energy profile strongly depends on how the domains are interacted. This finding has an important implication on the charge-boost induced by stabilization of negative capacitance in an FE + dielectric (DE) stack since the so-called capacitance matching needs to be designed at a specific operation voltage or frequency. In addition, we extract the domain viscosity dynamics during polarization switching according to the experimental measurements. For the first time, a physics-based circuit-compatible SPICE model for multi-domain phase field simulations is established to reveal the effect of domain interaction on the FE energy profile and microscopic domain evolution.

%\begin{description}
%\end{description}

\end{abstract}

\pacs{Valid PACS appear here}% PACS, the Physics and Astronomy
\maketitle

%\tableofcontents

\section{\label{sec:level1}Introduction}
Since the discovery back in the 1920s \cite{Valasek1920}, FE materials have attracted significant research attention because of (i) the ability to switch their polarization by an externally applied voltage and (ii) spontaneous polarization under zero bias. These unique properties make ferroelectrics promising materials for emerging nanoelectronic devices. One of the most fundamental yet important applications of FE materials is in ferroelectric capacitors, which consist of an FE layer sandwiched between two conducting metal contacts. Such a device structure is believed to have many prospective applications in both high density non-volatile memories and neuromorphic computing, including FE random access memories (FeRAMs), FE tunnel junctions (FTJs), and FE field-effect transistors (FeFETs)~\cite{Salahuddin2008,Garcia2014,Hsu2018}. \par % Importance and applications of FE materials

In recent decades, the relentless pursuit of Moore's law comes to a bottleneck due to the fact that as the device dimensions shrink, the power density in circuits becomes a challenging concern~\cite{Moore1998}. Based on Boltzmann statistics, the minimum voltage required for conventional complementary metal-oxide-semiconductor (CMOS) transistors to achieve a tenfold increase in the channel current is limited to $k_BT\ln10/q = 60$ mV at room temperature. This subthreshold swing SS $\geq 60$ mV/decade is considered as the fundamental limit for conventional CMOS transistors. Replacing the conventional gate oxide with FE materials was theoretically proposed to overcome the thermodynamic limit imposed on the power dissipation based on the stabilization of the negative capacitance (NC) region in the double-well energy profile predicted by the Landau theory~\cite{Salahuddin2008}. In Ref.~\cite{Salahuddin2008}, the transient nature of NC can be stabilized by connecting a proper DE or semiconductor capacitance in series, which leads to an internal voltage boost. Since then, a considerable amount of research efforts has been put into a thorough understanding of both transient and stabilized NC effects~\cite{Khan2014,Gao2014,Appleby2014,Zubko2016,Hoffmann2016,Chang2017a}. In particular, recently discovered doped hafnium FE materials are intensively studied due to the CMOS process compatibility~\cite{Boescke2011,Mueller2012,Sharma2017}. \par

Among all the research efforts, transient negative capacitance was observed during polarization switching in an R-FE capacitor (RFEC) circuit~\cite{Khan2014,Hoffmann2016}. Such transient NC was considered to be a direct indication of the negative capacitance region during polarization switching from one state to the other. To further characterize switching dynamics of a \ch{HfZrO2} (HZO) capacitor, the transient responses of the voltage across an FE ($V_\text{FE}$) were well measured experimentally with various pulse amplitudes in an RFEC circuit~\cite{Kobayashi2016}. With the single domain Landau-Khalatnikov theory, the transient NC and its link to the curvature of the free energy profile are explained by the mismatched switching rate of the free charge provided by the external circuit and the oxide bound charge~\cite{Chang2018}. From a multi-domain perspective, physics-based phase field models highlighted the importance of spatially-distributed FE grains and domain interaction in pulse switching dynamics~\cite{Yuan2016,Hoffmann2018,Saha2019}. Alternatively, the observed voltage drop was explained with conventional domain-mediated FE switching mechanisms based on the Kolmogorov-Avrami-Ishibashi (KAI) theory of domain nucleation and growth~\cite{Ishibashi1971,Kim2017}. \par

Despite all theoretical efforts to explain the experimental measurements of FE switching kinetics, quantitative frameworks that explore the voltage-dependent dynamic responses of ferroelectrics and the physical pictures of domain interaction based on experiments are still elusive. Therefore, it is critical to establish a physics-based theoretical model that can be calibrated with experimental measurements for the purpose of studying the multi-domain nature of ferroelectrics. To elucidate how microscopic domain interaction can change the macroscopic transient NC behaviors, in this work, we adopt a three-dimensional (3D) multi-domain phase field approach that can describe measured domain switching dynamics with well-calibrated parameters. In addition, current FE circuit models in the literature are developed only in either 1 or 2 dimensional space, and the effect of domain interaction is not considered in those models~\cite{Aziz2016,Hoffmann2016,Asai2017}. For the first time, we develop a physics-based circuit-compatible SPICE model that self-consistently performs 3D phase field simulations to further investigate multi-domain FE characteristics in an RFEC circuit. With this approach, we find that the polarization switching under different voltage pulses and the corresponding transient NC behaviors. By analyzing the effects of domain interaction on the free energy profile, it is shown that the free energy curvature strongly depends on the applied voltage. This finding implies that the depolarization-driven charge-boost realized in an FE/DE stack \cite{Chang2017a} needs to be designed under a specific voltage or frequency. Moreover, we show that the effect of domain interaction cannot be simply viewed as local effective electric field based on the fact that the gradient free energy significantly affects the total free energy landscape. By further calibrating the transient $V_\text{FE}$ with experiments, we obtain dynamic domain viscosity responses in the pulse measurements. \par

This paper is organized as follows. In Sec.~\ref{sec2}, a general phase field formalism is presented to describe the polarization switching dynamics of multi-domain FEs. Based on the phase field formulations described in Sec.~\ref{sec2}, a circuit compatible model is developed and implemented with the SPICE simulator in Sec.~\ref{sec3}. In Sec.~\ref{sec4}, the simulations results reveals the effects of domain interaction on HZO switching dynamics and the free energy profile. In Sec.~\ref{sec5}, we conclude this work by highlighting the multi-domain nature of HZO polarization switching and the importance of SPICE model implementation.

\section{\label{sec:level1}Phase Field Formalism} \label{sec2}
In order to capture the polycrystalline nature of HZO thin films, we adopt a comprehensive phase field framework that takes into account the contributions from the bulk free energy, gradient free energy, electric free energy and elastic free energy. Under this physics-based framework, the order parameter is a 3-dimensional polarization vector field $\vb{P} \qty(\vb{r},t)=\qty(P_1,P_2,P_3)$ as a function of space $\vb{r}=\qty(x_1,x_2,x_3)$ and time $t$. The temporal evolution of $\vb{P}$ is described by the time-dependent Ginzburg-Landau (TDGL) equation,
\begin{equation} \label{eq:TDGL}
\begin{split}
\pdv{P_i (\vb{r},t)}{t} &= -L\fdv{F\qty(P_i,\nabla P_i)}{P_i (\vb{r},t)} \\
&=-L\qty[\pdv{\qty(f_\text{bulk}+f_\text{elec}+f_\text{elas})}{P_i} - \nabla\cdot\pdv{f_\text{G}}{\nabla P_i}],
\end{split}
\end{equation}
where $L$ is the kinetic coefficient (inversely proportional to domain viscosity), and $i = 1,2,3$ stands for $x$, $y$ and $z$ directions, respectively~\cite{Landau1937,Ginzburg1945,Devonshire1949,Hong2008,Li2002a}. In general, the total free energy functional $F$ includes the bulk Landau free energy $f_\text{bulk}$, the gradient free energy $f_\text{G}$, the electric free energy $f_\text{elec}$, and the elastic free energy $f_\text{elas}$ over the film volume $V$: $F = \int \qty(f_\text{L} + f_\text{G} + f_\text{elec} + f_\text{elas}) \dd{V}$. However, due to the lack of discussion of elastic conditions in HZO thin films in the literature, the elastic energy contributions are excluded to highlight the effect of gradient free energy in this work. \par

The Landau free energy density can be expanded as
\begin{multline}
f_\text{bulk} = \alpha_1\qty(P_1^2 + P_2^2 + P_3^2) + \alpha_{11}\qty(P_1^4 + P_2^4 + P_3^4) \\
+ \alpha_{12}\qty(P_1^2P_2^2 + P_2^2P_3^2 + P_1^2P_3^2)  + \alpha_{111}\qty(P_1^6 + P_2^6 + P_3^6) \\
\alpha_{112}\qty[P_1^4\qty(P_2^2+P_3^2) + P_2^4\qty(P_1^2+P_3^2) + P_3^4\qty(P_1^2+P_2^2)] \\
+ \alpha_{123}\qty(P_1^2 P_2^2 P_3^2).
\end{multline}
where $\{\alpha_{i}\}$, $\{\alpha_{ij}\}$ and $\{\alpha_{ijk}\}$ are Landau expansion coefficients~\cite{Nambu1994}.

The gradient free energy results from the spatial gradients of the polarization and thus can be expressed as
\begin{equation}
\begin{split}
f_\text{G}&(P_{i,j}) = \frac{1}{2}G_{11}\qty(P^2_{1,1}+P^2_{2,2}+P^2_{3,3}) +\\
& \frac{1}{2}G_{44}\qty[2 P^2_{1,2}+2 P_{2,1}^2+2 P^2_{2,3}+2 P^2_{3,2}+2 P^2_{1,3}+2 P^2_{3,1}] \\
&= \frac{1}{2}G_{11}\left( P^2_{1,1}+P^2_{2,2}+P^2_{3,3} +  P^2_{1,2}+P^2_{2,1}+P^2_{2,3}  \right.\\
&\left. +P^2_{3,2}+P^2_{1,3}+P^2_{3,1} \right),
\end{split}
\end{equation}
where $P_{i,j} = \pdv{P_i}{x_j}$ is the spatial derivative of $\vb{P}$, $\{G_{ij}\}$ are the gradient coefficients that account for the FE domain interaction, and the second equality comes from $G_{44} = G_{11}/2$~\cite{Hong2008}.

The electric energy density can be calculated given the electric field and polarization as follows~\cite{Li2002a},
\begin{equation} \label{eq:f_elec}
f_\text{elec} = -\frac{1}{2}E_i \qty(\epsilon_0\kappa E_i + P_i).
\end{equation}
In Eq.~\eqref{eq:f_elec}, $\epsilon_0$ is the vacuum permittivity, $\vb{E}$ is the total electric field in the FE, $\kappa$ is the background dielectric constant that accounts for non-ferroelectric switching charges, and $\vb{P}$ is the ferroelectric-contributed polarization obtained from the TDGL equation~\cite{Tagantsev2008,Agarwal2019}. The electric displacement field $D_i = \epsilon_0\kappa E_i + P_i$ satisfies the electrostatic equation $D_{i,i} =0$ if there are no space charges inside the film~\cite{Li2002a}. The Poisson's equation (Eq.~\eqref{eq:poisson}) is obtained with $\vb{E}$ replaced by the potential gradient $-\nabla\phi$.
\begin{equation} \label{eq:poisson}
\epsilon_0\kappa\qty(\phi_{,11}+\phi_{,22}+\phi_{,33}) = \qty(P_{1,1} + P_{2,2} + P_{3,3}),
\end{equation}
which is subject to the out-of-plane boundary conditions (BCs):
\begin{equation}
\begin{split}
&\phi(z=0) = V_\text{FE}, \\
& \phi(z=t_\text{FE}) = 0.
\end{split}
\end{equation}
The in-plane boundary conditions are assumed to be periodic due to the fact that top metal contacts are patterned on a continuous FE thin film in experiments~\cite{Wang2006,Hong2008}. \par

With the aforementioned energy contributions, the governing TDGL equation can be numerically solved by the SPICE simulator with periodic BCs in the in-plane directions and zero BCs in the out-of-plane direction~\cite{Li2002a,Wang2006,Hong2008}.

\section{\label{sec:level1}SPICE Circuit model} \label{sec3}
In this section, we develop equivalent circuits of the TDGL equation and Poisson's equation so as to solve for the polarization and potential distributions self-consistently with the SPICE simulator.
\subsection{\label{sec:level2}TDGL equation}
The mathematical form of Eq. \eqref{eq:TDGL} can be viewed as an analogy to the voltage-current relationship of a unit capacitor $(C=1)$ \cite{Dutta2014}. In other words,
\begin{equation}
\dv{V(P_i)}{t} = I(V(P_1),V(P_2),V(P_3),V_{FE}) \Leftrightarrow C\dv{V}{t} = I,
\end{equation}
where $V(P_i),\ i=1,2,3$ is the voltage node for polarization in the SPICE simulator, and $I$ is a voltage-controlled current source as a function of polarization and FE voltage. It is noteworthy that the gradient energy contributions in the right hand side of Eq.~\eqref{eq:TDGL} can be simplified as
\begin{equation}
\begin{split}
\nabla\vdot\pdv{f_\text{G}}{\nabla P_i} &= \qty(\pdv{}{x},\pdv{}{y},\pdv{}{z})\vdot\qty(\pdv{f_\text{G}}{P_{i,1}},\pdv{f_\text{G}}{P_{i,2}},\pdv{f_\text{G}}{P_{i,3}}) \\
& = \pdv{}{x}\qty(\pdv{f_\text{G}}{P_{i,1}}) + \pdv{}{y}\qty(\pdv{f_\text{G}}{P_{i,2}}) + \pdv{}{z}\qty(\pdv{f_\text{G}}{P_{i,3}}) \\
& = G_{11}\qty(P_{i,11}+P_{i,22} + P_{i,33}) = G_{11}\nabla^2 P_i.
\end{split}
\end{equation}

With the finite difference discretization, the Laplacian of a variable can be expressed as
\begin{equation}
\begin{split}
\nabla^2 P_i = &\frac{P_i(m+1,n,k) + P_i(m-1,n,k) - 2P_i(m,n,k)}{dx^2}\\
&+ \frac{P_i(m,n+1,k) + P_i(m,n-1,k) - 2P_i(m,n,k)}{dy^2}\\
&+ \frac{P_i(m,n,k+1) + P_i(m,n,k-1) - 2P_i(m,n,k)}{dz^2},
\end{split}
\end{equation}
where $\{dx,dy,dz\}$ are the numerical grid spacing and $\{m,n,k\}$ are discrete indices in each dimension.

\begin{figure}[!t]
\centering
\includegraphics[width=0.45\textwidth]{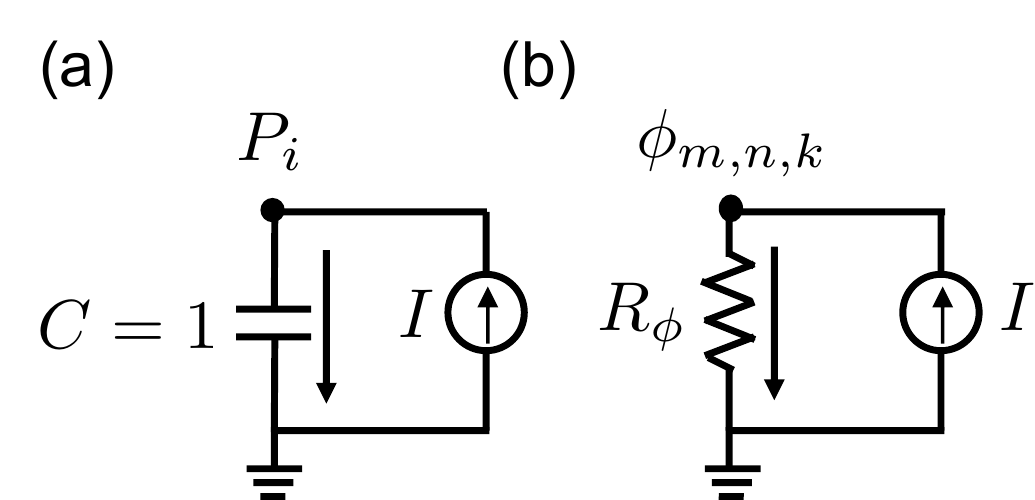}
\caption{The SPICE equivalent circuit diagrams of (a) the TDGL equation, where $i=1,2,3$ and (b) the Poisson's equation.}
\label{fig:equi_ckts}
\end{figure}

\subsection{\label{sec:level2}Poisson's equation}
To obtain the multi-domain potential profile and corresponding local electric field, we discretize Eq.~\eqref{eq:poisson} as
\begin{equation} \label{eq:10}
\begin{split}
\nabla^2\phi = &\frac{\phi_{m+1,n,k} + \phi_{m-1,n,k} - 2\phi_{m,n,k}}{dx^2} \\
&+ \frac{\phi_{m,n+1,k} + \phi_{m,n-1,k} - 2\phi_{m,n,k}}{dy^2} \\
&+ \frac{\phi_{m,n,k+1} + \phi_{m,n,k-1} - 2\phi_{m,n,k}}{dz^2} = g(\vb{P}),
\end{split}
\end{equation}
where $g(\vb{P})=\qty(P_{1,1}+P_{2,2}+P_{3,3})/\qty(\epsilon_0\kappa)$. By rearranging Eq.~\eqref{eq:10}, one obtains
\begin{equation}\label{eq:11}
\begin{split}
\frac{\phi_{m,n,k}}{R_\phi} &= -g(\vb{P}) + \frac{\phi_{m+1,n,k} + \phi_{m-1,n,k}}{dx^2} \\
&\quad + \frac{\phi_{m,n+1,k} + \phi_{m,n-1,k}}{dy^2} + \frac{\phi_{m,n,k+1} + \phi_{m,n,k-1}}{dz^2}\\
&= I\qty(\vb{P},\phi_{nn}),
\end{split}
\end{equation}
where $R_\phi=1/\qty(2/dx^2+2/dy^2+2/dz^2)$ and $\phi_{nn}$ represents the potentials of the nearest neighboring cells. Eq.~\eqref{eq:11} can be viewed as the voltage-current relationship of a constant resistor $R_\phi$ with the right hand side being a voltage-controlled current source.

\begin{figure}[!t]
\centering
\includegraphics[width=0.3\textwidth]{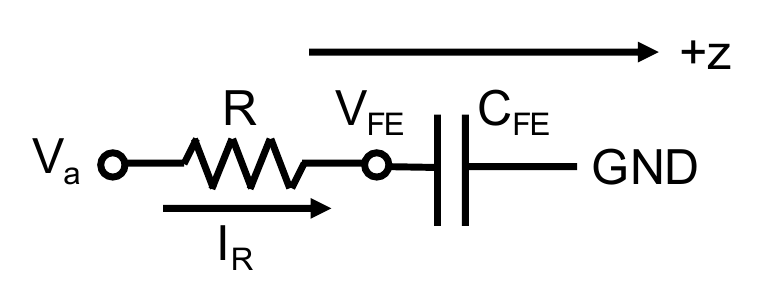}
\caption{The RFEC circuit diagram used in Ref.~\cite{Kobayashi2016} and this work.}
\label{fig:rfec}
\end{figure}

If the finite screening lengths of the metal contacts are considered, the potential boundary conditions will depend on out-of-plane polarization as follows in order to account for the depolarization effect~\cite{Chang2017}.
\begin{equation}
\begin{split}
&V(z=0) = V_\text{FE} - \frac{\sigma_s\lambda_1}{\epsilon_1\epsilon_0}\\
&V(z=t_\text{FE}) = 0 + \frac{\sigma_s\lambda_2}{\epsilon_2\epsilon_0},
\end{split}
\end{equation}
where $\sigma_s=\frac{P_z t_\text{FE}}{t_\text{FE}+\lambda_1/\epsilon_1 + \lambda_2/\epsilon_2}$ is the screening charge density due to out-of-plane polarization, and $\lambda_{1,2}$ and $\epsilon_{1,2}$ are the screening lengths and relative dielectric constants of contacts \{1,2\}, respectively.
The equivalent circuit diagrams of Eq.~\eqref{eq:TDGL} and Eq.~\eqref{eq:poisson} are summarized in Fig.~\ref{fig:equi_ckts}. Note that it is important to distinguish the total free charge density and polarization in an RFEC circuit~\cite{Chang2018}. In Fig.~\ref{fig:rfec}, the current flowing through the FE capacitor $I_R$ can be calculated as
\begin{equation} \label{eq:I_R}
I_R= \dv{Q_\text{FE}}{t} = \dv{}{t}\qty(\epsilon_0\kappa \frac{V_\text{FE}}{t_\text{FE}} + P_z)A,
\end{equation}
where $Q_\text{FE}$ is the total free charge, $A$ is the capacitor cross-sectional area and $P_z$ is the average polarization in the out-of-plane direction. Note that it is more convenient to implement the contact module with Verilog-A based on Eq.~\eqref{eq:I_R}. It is noteworthy that the measured charge is the total free charge $Q_\text{FE}$ instead of polarization charge according to Eq.~\eqref{eq:I_R}.

\section{\label{sec:level1}Results and Discussion} \label{sec4}
Based on the phase field formalism, we solve the TDGL equation (Eq.~\eqref{eq:TDGL}) and Poisson's equation (Eq.~\eqref{eq:poisson}) for polarization charge and potential distributions to investigate the pulse switching dynamics of multi-domain HZO capacitors in an RFEC circuit. In this work, the baseline experimental measurements are extracted from Ref.~\cite{Kobayashi2016}, which demonstrated the transient responses of a TiN/\ch{Hf_{0.7}Zr_{0.3}O2}/TiN capacitor under various pulse amplitudes. The parameters used in this work are summarized in Table~\ref{tab:1}.

%% table
\begin{table}[!t]  %% increase table row spacing, adjust to taste
\renewcommand{\arraystretch}{1.3}  % if using array.sty, it might be a good idea to tweak the value of
% \extrarowheight as needed to properly center the text within the cells.
%% Some packages, such as MDW tools, offer better commands for making tables
%% than the plain LaTeX2e tabular which is used here.
\centering
\caption{Parameters Used For HZO in This Work} \label{tab:1}
\begin{threeparttable}
\begin{tabular}{c c}
\hline \hline
Parameter & Value\\ \hline
$\qty(\lambda/\epsilon)_{1,2}$ (\AA) & 0.06,\ 0.06 \\ \hline
$\kappa$ & 35 \\ \hline
$t_\text{FE}$ \si{(nm)} & $10$~\cite{Kobayashi2016} \\ \hline
$\alpha_1$ (m/F) & \num{-4e8}~\cite{Chang2018} \\ \hline
$\alpha_{11}$ \si{(m^5 C^{-2}/F)}  & \num{3.7e9}~\cite{Chang2018} \\ \hline
$\alpha_{111}$ \si{(m^9 C^{-4}/F)} & \num{1.1e9}~\cite{Chang2018} \\ \hline
$G_{11}$ \si{(m^3/F)} & \textit{dynamic}\\ \hline
$L$ \si{(\ohm m)^{-1}} & $\num{2e-4}$ or \textit{dynamic} \\ \hline
$R$ \si{(\ohm)} & 20k~\cite{Kobayashi2016} \\ \hline
Area \si{(m^2)} & $\num{7e-9}$~\cite{Kobayashi2016} \\ \hline \hline
%$W_p$\tnote{1} (\si{nm}) & 180 \\ \hline \hline
\end{tabular}
%\begin{tablenotes}
%%   \item This is the caption.
%	\item[1] $Q_{ij}$ and $c_{ij}$ of HZO are not well reported in the literature. Here, we use values for other FE materials instead.
%	\item[2] $R_1$ and $R_2$ are series resistance for hysteresis and pulse switching simulations, respectively.
%\end{tablenotes}

\end{threeparttable}
\end{table}

\begin{figure}[!t]
\centering
\includegraphics[width=0.45\textwidth]{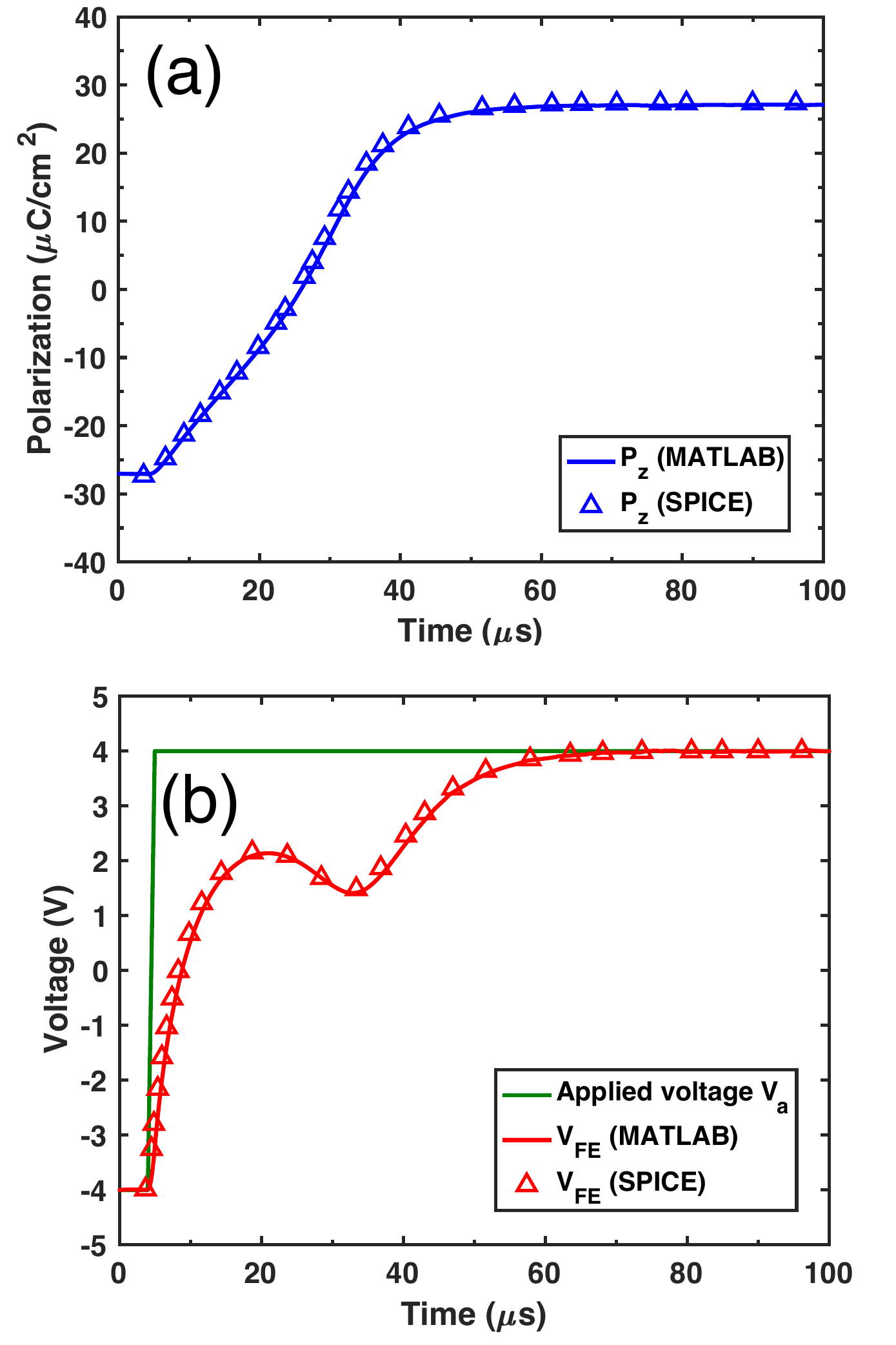}
\caption{The comparisons between the semi-implicit Fourier-spectral method and SPICE circuit simulations of $P_\text{z}$ and $V_\text{FE}$ with $G_{11}=\SI{1e-9}{(m^3/F)}$ at a pulse amplitude of $\SI{4}{V}$.}
\label{fig:spice_matlab}
\end{figure}

\begin{figure*}[!t]
\centering
\includegraphics[width=1\textwidth]{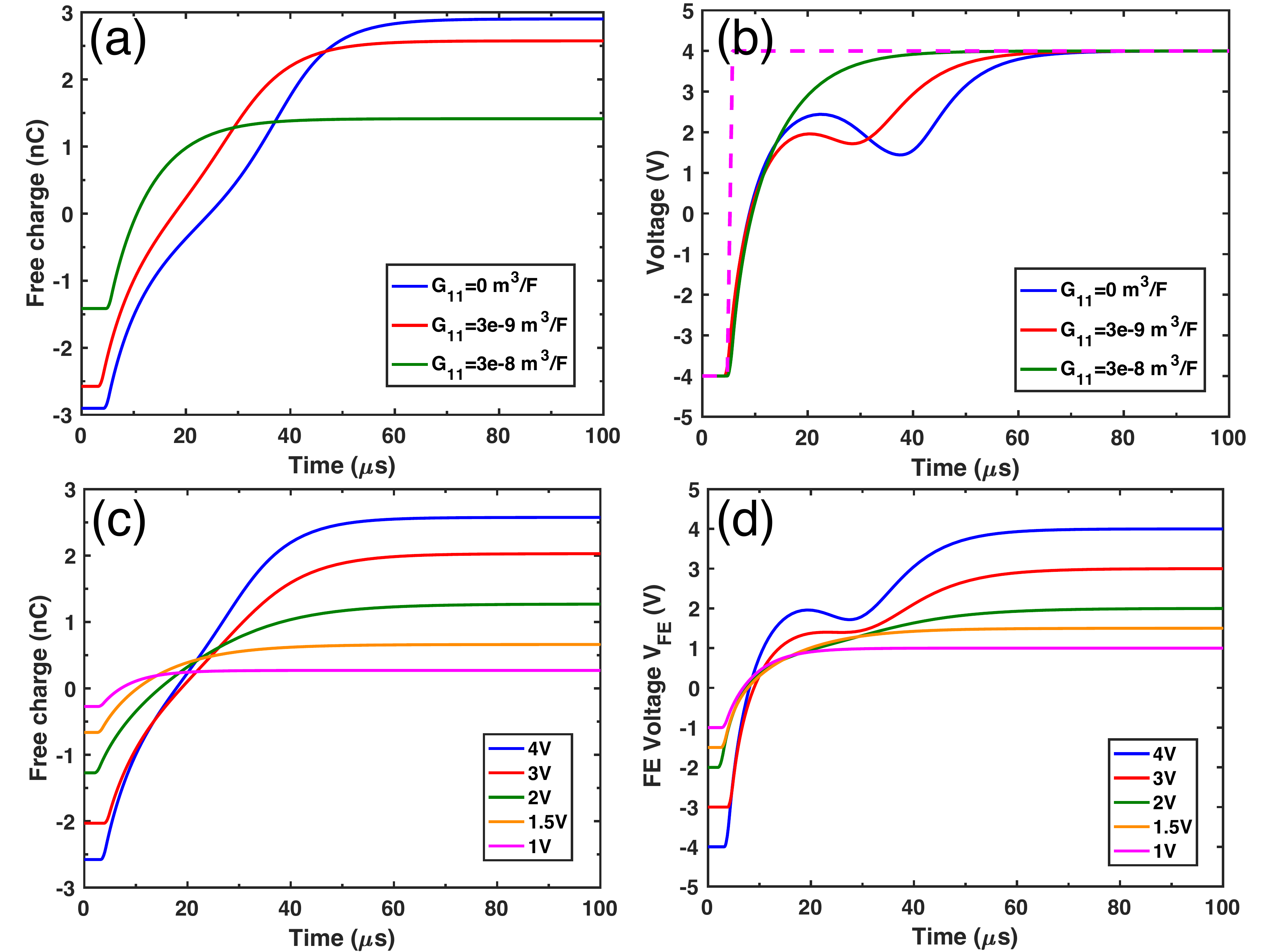}
\caption{The SPICE simulations of (a) charge switching dynamics and (b) $V_\text{FE}$ transient responses with various $G_{11}$ values under a voltage pulse of $\SI{4}{V}$ (dashed line). (c) and (d) show transient responses of $Q_\text{FE}$ and $V_\text{FE}$ with $G_{11}$ correction under various pulse amplitudes.}
\label{fig:tran_g11}
\end{figure*}

\subsection{\label{sec:level2}SPICE model initialization and validation}
In our simulations, we first obtain the steady state multi-domain polarization distributions and potential profile under zero bias with polarization initialized with a zero-mean normal distribution. A negative voltage pulse is applied to the steady-state domain state to obtain the initial conditions for the pulse measurements. \par

To validate the proposed SPICE model, we also solve the TDGL equation (Eq.~\eqref{eq:TDGL}) and Poisson's equation (Eq.~\eqref{eq:poisson}) in MATLAB using the semi-implicit Fourier-spectral method~\cite{Chen1998}. Fig.~\ref{fig:spice_matlab} shows the simulated transient responses from the SPICE simulator compared to those from the Fourier-spectral method with the same numerical settings and boundary conditions. For simplicity, the parasitic capacitance parallel to the FE capacitor is assumed to be small enough to be ignored in the RFEC circuit~\cite{Hoffmann2016}. The simulations presented below are all from the SPICE simulator if not mentioned elsewhere.

\begin{figure*}[!t]
\centering
\includegraphics[width=1\textwidth]{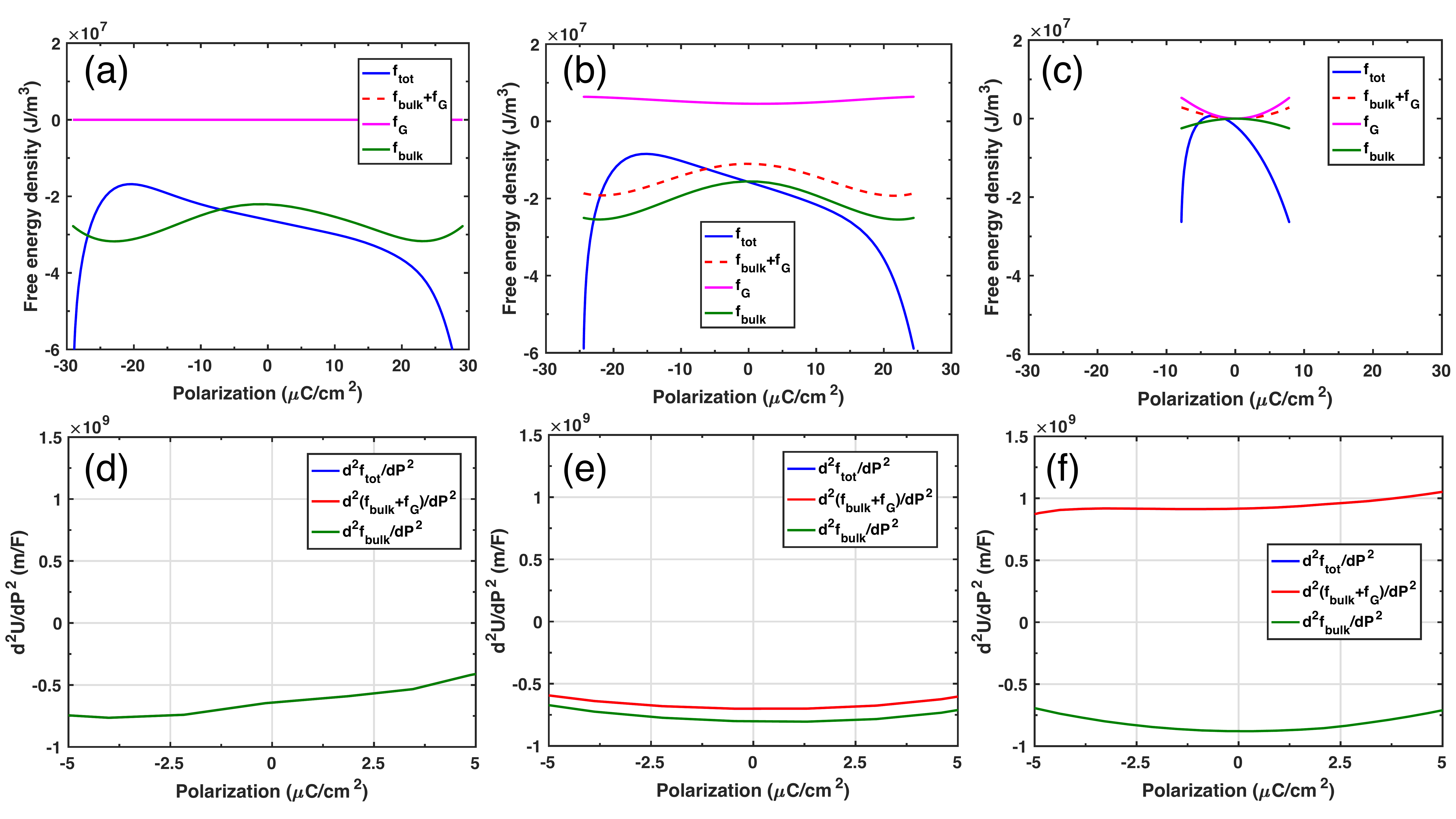}
\caption{The top row shows transient energy profiles during pulse switching at various $G_{11}$. The corresponding free energy curvatures are in the bottom row. The $G_{11}$ values for \{(a)(d)\}, \{(b)(e)\} and \{(c)(f)\} are $\SI{0}{m^3/F}$, $\SI{3e-9}{m^3/F}$ and $\SI{3e-8}{m^3/F}$, respectively. The total free energy $f_\text{tot} = f_\text{bulk}+f_\text{G}+f_\text{elec}$.}
\label{fig:energy}
\end{figure*}

\subsection{\label{sec:level2}Effects of $G_{11}$ on the transient responses}
\begin{figure*}[!t]
\centering
\includegraphics[width=1\textwidth]{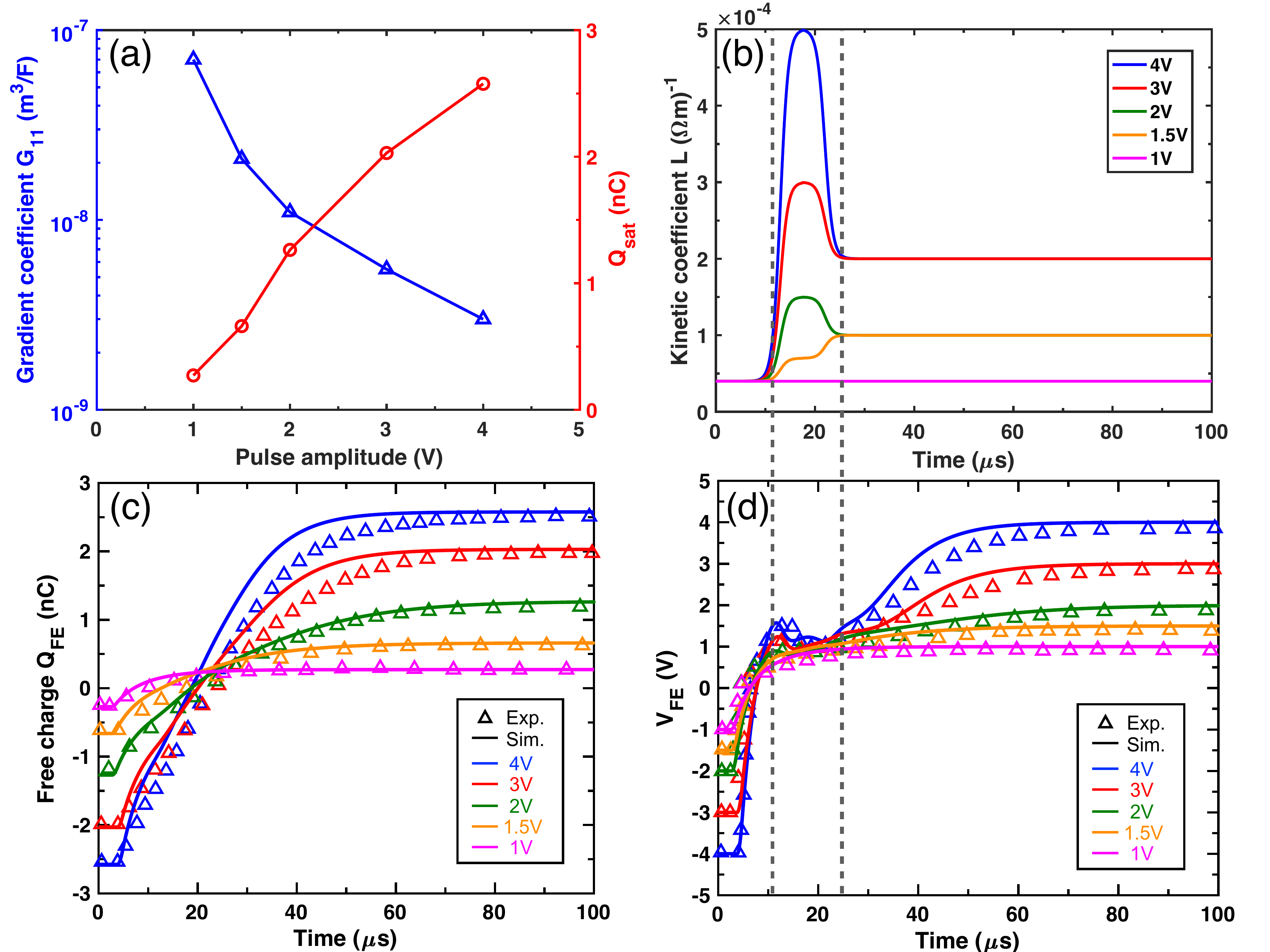}
\caption{(a) The extracted $G_{11}$ values and the corresponding saturation free charge $Q_\text{sat}$ at various pulse amplitudes according to the experimental measurements in Ref.~\cite{Kobayashi2016}. (b) Kinetic coefficient dynamics during polarization switching at various pulse amplitudes. (c) The SPICE simulations of free charge switching dynamics at various pulse amplitudes compared with experiments in Ref.~\cite{Kobayashi2016}. (d) The SPICE simulations of $V_\text{FE}$ transient responses at various pulse amplitudes compared with experiments in Ref.~\cite{Kobayashi2016}.}
\label{fig:sim_exp}
\end{figure*}

From the experimental measurements in Ref.~\cite{Kobayashi2016}, the charge in the steady state is found to be suppressed at smaller pulse amplitudes. With our model, we find that the measured saturation free charges at various pulse amplitudes do not match those predicted by the phase field approach with a constant $G_{11}$. As a result, we propose that the gradient coefficient $G_{11}$ (and hence domain interaction) depends on the applied voltage. To verify this argument, we first examine how $G_{11}$ affects the transient behaviors at a pulse amplitude of $\SI{4}{V}$. In Fig.~\ref{fig:tran_g11}(a), the saturation charge decreases significantly as $G_{11}$ increases, which indicates that stronger domain interaction tends to suppress polarization switching. Moreover, the experimental measurements in Ref.~\cite{Kobayashi2016} also show that the transient voltage drop disappears as the pulse voltage decreases. This finding is also captured by increasing $G_{11}$. As shown in Fig.~\ref{fig:tran_g11}(b), $V_\text{FE}$ dynamics acts like a normal dielectric capacitor and the voltage drop is no longer to be seen as domain interaction gets stronger (or larger $G_{11}$). Next, we extract the $G_{11}$ value for each pulse amplitude based on the saturation total free charge measured in Ref.~\cite{Kobayashi2016}. Fig.~\ref{fig:sim_exp}(a) shows the extracted $G_{11}$ at each pulse amplitude based on the experimentally measured FE free charge in the steady state. With extracted $G_{11}$, Fig.~\ref{fig:tran_g11}(c) and (d) show the transient responses of $Q_\text{FE}$ and $V_\text{FE}$ at various pulse amplitudes, which are qualitatively consistent with the experimental observations. Therefore, these simulation results imply that domain interaction of the HZO thin film is weaker under a larger voltage, which may be attributed to the breaking of spatial domain coupling under a large electric field.

\subsection{\label{sec:level2}Effects of $G_{11}$ on the free energy profile}
Now that we have shown that $G_{11}$ is dependent on the applied voltage, we investigate the effects of domain interaction on the observed transient responses in terms of the free energy profile. Fig.~\ref{fig:energy}(a)--(c) show the transient energy profiles when polarization switches from the negative state to the positive state under a pulse of $\SI{4}{V}$ with increasing $G_{11}$. Without domain interaction ($G_{11}=\SI{0}{m^3/F}$), the bulk free energy exhibits a negative capacitance region (negative curvature) during polarization switching. With a nonzero $G_{11}$, the gradient free energy due to FE domain interaction shows a quadratic shape, and thus turns the negative curvature of the bulk free energy into a positive one as $G_{11}$ increases. Note that the extracted gradient free energy as a function of polarization is consistent with the parabolic shape obtained from the first-principles calculations~\cite{Li2017}. The corresponding free energy curvatures $\qty(\pdv[2]{U}{P_z})$ near zero polarization are plotted in Fig.~\ref{fig:energy}(d)--(f). The positive curvature of the total free energy with a larger $G_{11}$ explains how stronger domain interaction can suppress the transient NC. Since $G_{11}$ is voltage-dependent, the energy profile at various voltages will depend on how FE domains are interacted.\par

The voltage-dependence of the free energy curvature implies that FE capacitance can only be matched with a constant DE capacitance for charge-boost at a specific voltage based on the fact that the total free energy curvature is directly related to the FE capacitance~\cite{Khan2014,Chang2018}. Moreover, in Fig.~\ref{fig:tran_g11}(c) and (d), we use a constant value of $G_{11}$ to simulate the switching dynamics at an applied voltage pulse due to the short rise time of the pulse. In reality, $G_{11}$ varies when the applied voltage switches from low to high. This indicates that the frequency of the applied pulse affects the FE-DE capacitance matching as well.\par

Note that the electric free energy has a mathematical form as in Eq.~\eqref{eq:f_elec} and hence does not affect the energy curvature, as can be seen in Fig.~\ref{fig:energy}(d)--(f). As a result, our simulations also show that the gradient energy contribution $G_{11}\nabla^2P_i$ cannot be simply treated as an effective interaction field because the existence of gradient free energy changes the curvature of the total free energy landscape~\cite{Saha2019}.

\subsection{\label{sec:level2}Dynamic kinetic coefficient}
With extracted $G_{11}$ and a constant kinetic coefficient $L$, the measured saturation polarization can be well captured by the phase field framework. However, the voltage responses during the transient NC are not consistent with experiments. Therefore, we adopt a dynamic $L$ to further characterize the domain viscosity variations during polarization switching. Fig.~\ref{fig:sim_exp}(b) shows the transient kinetic coefficient $L$ at various pulse amplitudes. When $V_\text{FE}$ is below the coercive voltage $V_c$, the constant $L$ represents the inherent dielectric response. The larger $L$ in the time interval between the gray dashed lines demonstrates that the polarization switching speeds up due to the unstable nature of the NC region. As the FE capacitance goes back to a positive value, $L$ decreases and the transient $V_\text{FE}$ recovers to a normal dielectric response. For smaller pulse amplitudes, the energy curvature is positive during polarization switching due to stronger domain interaction, and therefore the transient responses of the FE are similar to DE responses without a voltage drop. Compared with measurements in Ref.~\cite{Kobayashi2016}, the simulation results of total free charge $Q_\text{FE}$ and FE voltage $V_\text{FE}$ with various pulse amplitudes are shown in Fig.~\ref{fig:sim_exp}(c) and (d).\par
Although our simulations show reasonable trends of the measured transient responses, there are still some discrepancies between the simulations and experiments, which may result from the elastic free energy. Because HZO thin films are not classical perovskites, the crystal structures and the related elastic contributions to the total free energy need further experimental investigations, which is beyond the scope of this paper. Further simulations that include broad distributions of bulk Landau parameters show insignificant impacts on the transient responses and confirm the dominant effects of the gradient coefficient~\cite{supplemental_material}. However, the gradient coefficient and kinetic coefficient may also have spatial distributions due to the multi-domain nature of HZO~\cite{Hoffmann2016,Hoffmann2018}.

\section{\label{sec:level1}Conclusion} \label{sec5}
In summary, the first physics-based circuit compatible SPICE model for multi-domain ferroelectric materials is developed and calibrated with experimental measurements. With this model, we investigate the effect of domain interaction on the transient responses of HZO capacitors under a voltage pulse in an RFEC circuit. We find that FE domain interaction depends on the applied voltage and plays an important role in the dynamic responses of polarization switching and the transient NC effect. By studying how domain interaction affects the total free energy curvature, we show that the effect of domain interaction cannot be viewed as an effective electric field. More importantly, the voltage-dependent domain interaction indicates that FE-DE capacitance matching  can only be achieved at a specific voltage and frequency. Furthermore, the dynamic nature of FE domain viscosity at a voltage pulse is explored based on the experimental measurements. This work explores the physical roles that phenomenological parameters play in microscopic switching mechanisms for multi-domain HZO capacitors, and the proposed circuit model shows the potential for the analyses of \ch{HfO2}-based ferroelectrics at device and circuit levels.

%\section*{\label{sec:level1}Acknowledgments}
%This work was funded by Intel Corporation through Semiconductor Research Corporation MSR-INTEL TASK 2835.001.

\RaggedRight
\bibliography{./PRApplied}

\begin{thebibliography}{37}%
\makeatletter
\providecommand \@ifxundefined [1]{%
 \@ifx{#1\undefined}
}%
\providecommand \@ifnum [1]{%
 \ifnum #1\expandafter \@firstoftwo
 \else \expandafter \@secondoftwo
 \fi
}%
\providecommand \@ifx [1]{%
 \ifx #1\expandafter \@firstoftwo
 \else \expandafter \@secondoftwo
 \fi
}%
\providecommand \natexlab [1]{#1}%
\providecommand \enquote  [1]{``#1''}%
\providecommand \bibnamefont  [1]{#1}%
\providecommand \bibfnamefont [1]{#1}%
\providecommand \citenamefont [1]{#1}%
\providecommand \href@noop [0]{\@secondoftwo}%
\providecommand \href [0]{\begingroup \@sanitize@url \@href}%
\providecommand \@href[1]{\@@startlink{#1}\@@href}%
\providecommand \@@href[1]{\endgroup#1\@@endlink}%
\providecommand \@sanitize@url [0]{\catcode `\\12\catcode `\$12\catcode
  `\&12\catcode `\#12\catcode `\^12\catcode `\_12\catcode `\%12\relax}%
\providecommand \@@startlink[1]{}%
\providecommand \@@endlink[0]{}%
\providecommand \url  [0]{\begingroup\@sanitize@url \@url }%
\providecommand \@url [1]{\endgroup\@href {#1}{\urlprefix }}%
\providecommand \urlprefix  [0]{URL }%
\providecommand \Eprint [0]{\href }%
\providecommand \doibase [0]{http://dx.doi.org/}%
\providecommand \selectlanguage [0]{\@gobble}%
\providecommand \bibinfo  [0]{\@secondoftwo}%
\providecommand \bibfield  [0]{\@secondoftwo}%
\providecommand \translation [1]{[#1]}%
\providecommand \BibitemOpen [0]{}%
\providecommand \bibitemStop [0]{}%
\providecommand \bibitemNoStop [0]{.\EOS\space}%
\providecommand \EOS [0]{\spacefactor3000\relax}%
\providecommand \BibitemShut  [1]{\csname bibitem#1\endcsname}%
\let\auto@bib@innerbib\@empty
%</preamble>
\bibitem [{\citenamefont {Valasek}(1920)}]{Valasek1920}%
  \BibitemOpen
  \bibfield  {author} {\bibinfo {author} {\bibfnamefont {J.}~\bibnamefont
  {Valasek}},\ }\href@noop {} {\bibfield  {journal} {\bibinfo  {journal} {MSc
  thesis, Univ. Minnesota}\ } (\bibinfo {year} {1920})}\BibitemShut {NoStop}%
\bibitem [{\citenamefont {Salahuddin}\ and\ \citenamefont
  {Datta}(2008)}]{Salahuddin2008}%
  \BibitemOpen
  \bibfield  {author} {\bibinfo {author} {\bibfnamefont {S.}~\bibnamefont
  {Salahuddin}}\ and\ \bibinfo {author} {\bibfnamefont {S.}~\bibnamefont
  {Datta}},\ }\href {http://dx.doi.org/10.1021/nl071804g} {\bibfield  {journal}
  {\bibinfo  {journal} {Nano Letters}\ }\textbf {\bibinfo {volume} {8}},\
  \bibinfo {pages} {405} (\bibinfo {year} {2008})},\ \bibinfo {note} {doi:
  \url{10.1021/nl071804g}},\ \Eprint
  {http://arxiv.org/abs/http://dx.doi.org/10.1021/nl071804g}
  {http://dx.doi.org/10.1021/nl071804g} \BibitemShut {NoStop}%
\bibitem [{\citenamefont {Garcia}\ and\ \citenamefont
  {Bibes}(2014)}]{Garcia2014}%
  \BibitemOpen
  \bibfield  {author} {\bibinfo {author} {\bibfnamefont {V.}~\bibnamefont
  {Garcia}}\ and\ \bibinfo {author} {\bibfnamefont {M.}~\bibnamefont {Bibes}},\
  }\href@noop {} {\bibfield  {journal} {\bibinfo  {journal} {Nature
  Communications}\ } (\bibinfo {year} {2014})}\BibitemShut {NoStop}%
\bibitem [{\citenamefont {Hsu}\ \emph {et~al.}(2018)\citenamefont {Hsu},
  \citenamefont {Pan},\ and\ \citenamefont {Naeemi}}]{Hsu2018}%
  \BibitemOpen
  \bibfield  {author} {\bibinfo {author} {\bibfnamefont {C.-S.}\ \bibnamefont
  {Hsu}}, \bibinfo {author} {\bibfnamefont {C.}~\bibnamefont {Pan}}, \ and\
  \bibinfo {author} {\bibfnamefont {A.}~\bibnamefont {Naeemi}},\ }\href
  {\doibase 10.1109/led.2018.2820118} {\bibfield  {journal} {\bibinfo
  {journal} {{IEEE} Electron Device Letters}\ }\textbf {\bibinfo {volume}
  {39}},\ \bibinfo {pages} {765} (\bibinfo {year} {2018})}\BibitemShut
  {NoStop}%
\bibitem [{\citenamefont {Moore}(1998)}]{Moore1998}%
  \BibitemOpen
  \bibfield  {author} {\bibinfo {author} {\bibfnamefont {G.}~\bibnamefont
  {Moore}},\ }\href {\doibase 10.1109/jproc.1998.658762} {\bibfield  {journal}
  {\bibinfo  {journal} {Proceedings of the {IEEE}}\ }\textbf {\bibinfo {volume}
  {86}},\ \bibinfo {pages} {82} (\bibinfo {year} {1998})}\BibitemShut {NoStop}%
\bibitem [{\citenamefont {Khan}\ \emph {et~al.}(2014)\citenamefont {Khan},
  \citenamefont {Chatterjee}, \citenamefont {Wang}, \citenamefont {Drapcho},
  \citenamefont {You}, \citenamefont {Serrao}, \citenamefont {Bakaul},
  \citenamefont {Ramesh},\ and\ \citenamefont {Salahuddin}}]{Khan2014}%
  \BibitemOpen
  \bibfield  {author} {\bibinfo {author} {\bibfnamefont {A.~I.}\ \bibnamefont
  {Khan}}, \bibinfo {author} {\bibfnamefont {K.}~\bibnamefont {Chatterjee}},
  \bibinfo {author} {\bibfnamefont {B.}~\bibnamefont {Wang}}, \bibinfo {author}
  {\bibfnamefont {S.}~\bibnamefont {Drapcho}}, \bibinfo {author} {\bibfnamefont
  {L.}~\bibnamefont {You}}, \bibinfo {author} {\bibfnamefont {C.}~\bibnamefont
  {Serrao}}, \bibinfo {author} {\bibfnamefont {S.~R.}\ \bibnamefont {Bakaul}},
  \bibinfo {author} {\bibfnamefont {R.}~\bibnamefont {Ramesh}}, \ and\ \bibinfo
  {author} {\bibfnamefont {S.}~\bibnamefont {Salahuddin}},\ }\href {\doibase
  10.1038/nmat4148} {\bibfield  {journal} {\bibinfo  {journal} {Nature
  Materials}\ }\textbf {\bibinfo {volume} {14}},\ \bibinfo {pages} {182}
  (\bibinfo {year} {2014})}\BibitemShut {NoStop}%
\bibitem [{\citenamefont {Gao}\ \emph {et~al.}(2014)\citenamefont {Gao},
  \citenamefont {Khan}, \citenamefont {Marti}, \citenamefont {Nelson},
  \citenamefont {Serrao}, \citenamefont {Ravichandran}, \citenamefont
  {Ramesh},\ and\ \citenamefont {Salahuddin}}]{Gao2014}%
  \BibitemOpen
  \bibfield  {author} {\bibinfo {author} {\bibfnamefont {W.}~\bibnamefont
  {Gao}}, \bibinfo {author} {\bibfnamefont {A.}~\bibnamefont {Khan}}, \bibinfo
  {author} {\bibfnamefont {X.}~\bibnamefont {Marti}}, \bibinfo {author}
  {\bibfnamefont {C.}~\bibnamefont {Nelson}}, \bibinfo {author} {\bibfnamefont
  {C.}~\bibnamefont {Serrao}}, \bibinfo {author} {\bibfnamefont
  {J.}~\bibnamefont {Ravichandran}}, \bibinfo {author} {\bibfnamefont
  {R.}~\bibnamefont {Ramesh}}, \ and\ \bibinfo {author} {\bibfnamefont
  {S.}~\bibnamefont {Salahuddin}},\ }\href {\doibase 10.1021/nl502691u}
  {\bibfield  {journal} {\bibinfo  {journal} {Nano Letters}\ }\textbf {\bibinfo
  {volume} {14}},\ \bibinfo {pages} {5814} (\bibinfo {year}
  {2014})}\BibitemShut {NoStop}%
\bibitem [{\citenamefont {Appleby}\ \emph {et~al.}(2014)\citenamefont
  {Appleby}, \citenamefont {Ponon}, \citenamefont {Kwa}, \citenamefont {Zou},
  \citenamefont {Petrov}, \citenamefont {Wang}, \citenamefont {Alford},\ and\
  \citenamefont {O'Neill}}]{Appleby2014}%
  \BibitemOpen
  \bibfield  {author} {\bibinfo {author} {\bibfnamefont {D.~J.~R.}\
  \bibnamefont {Appleby}}, \bibinfo {author} {\bibfnamefont {N.~K.}\
  \bibnamefont {Ponon}}, \bibinfo {author} {\bibfnamefont {K.~S.~K.}\
  \bibnamefont {Kwa}}, \bibinfo {author} {\bibfnamefont {B.}~\bibnamefont
  {Zou}}, \bibinfo {author} {\bibfnamefont {P.~K.}\ \bibnamefont {Petrov}},
  \bibinfo {author} {\bibfnamefont {T.}~\bibnamefont {Wang}}, \bibinfo {author}
  {\bibfnamefont {N.~M.}\ \bibnamefont {Alford}}, \ and\ \bibinfo {author}
  {\bibfnamefont {A.}~\bibnamefont {O'Neill}},\ }\href {\doibase
  10.1021/nl5017255} {\bibfield  {journal} {\bibinfo  {journal} {Nano Letters}\
  }\textbf {\bibinfo {volume} {14}},\ \bibinfo {pages} {3864} (\bibinfo {year}
  {2014})}\BibitemShut {NoStop}%
\bibitem [{\citenamefont {Zubko}\ \emph {et~al.}(2016)\citenamefont {Zubko},
  \citenamefont {Wojde{\l}}, \citenamefont {Hadjimichael}, \citenamefont
  {Fernandez-Pena}, \citenamefont {Sen{\'{e}}}, \citenamefont {Luk'yanchuk},
  \citenamefont {Triscone},\ and\ \citenamefont
  {{\'{I}}{\~{n}}iguez}}]{Zubko2016}%
  \BibitemOpen
  \bibfield  {author} {\bibinfo {author} {\bibfnamefont {P.}~\bibnamefont
  {Zubko}}, \bibinfo {author} {\bibfnamefont {J.~C.}\ \bibnamefont
  {Wojde{\l}}}, \bibinfo {author} {\bibfnamefont {M.}~\bibnamefont
  {Hadjimichael}}, \bibinfo {author} {\bibfnamefont {S.}~\bibnamefont
  {Fernandez-Pena}}, \bibinfo {author} {\bibfnamefont {A.}~\bibnamefont
  {Sen{\'{e}}}}, \bibinfo {author} {\bibfnamefont {I.}~\bibnamefont
  {Luk'yanchuk}}, \bibinfo {author} {\bibfnamefont {J.-M.}\ \bibnamefont
  {Triscone}}, \ and\ \bibinfo {author} {\bibfnamefont {J.}~\bibnamefont
  {{\'{I}}{\~{n}}iguez}},\ }\href {\doibase 10.1038/nature17659} {\bibfield
  {journal} {\bibinfo  {journal} {Nature}\ }\textbf {\bibinfo {volume} {534}},\
  \bibinfo {pages} {524} (\bibinfo {year} {2016})}\BibitemShut {NoStop}%
\bibitem [{\citenamefont {Hoffmann}\ \emph {et~al.}(2016)\citenamefont
  {Hoffmann}, \citenamefont {Pe{\v{s}}i{\'{c}}}, \citenamefont {Chatterjee},
  \citenamefont {Khan}, \citenamefont {Salahuddin}, \citenamefont {Slesazeck},
  \citenamefont {Schroeder},\ and\ \citenamefont {Mikolajick}}]{Hoffmann2016}%
  \BibitemOpen
  \bibfield  {author} {\bibinfo {author} {\bibfnamefont {M.}~\bibnamefont
  {Hoffmann}}, \bibinfo {author} {\bibfnamefont {M.}~\bibnamefont
  {Pe{\v{s}}i{\'{c}}}}, \bibinfo {author} {\bibfnamefont {K.}~\bibnamefont
  {Chatterjee}}, \bibinfo {author} {\bibfnamefont {A.~I.}\ \bibnamefont
  {Khan}}, \bibinfo {author} {\bibfnamefont {S.}~\bibnamefont {Salahuddin}},
  \bibinfo {author} {\bibfnamefont {S.}~\bibnamefont {Slesazeck}}, \bibinfo
  {author} {\bibfnamefont {U.}~\bibnamefont {Schroeder}}, \ and\ \bibinfo
  {author} {\bibfnamefont {T.}~\bibnamefont {Mikolajick}},\ }\href {\doibase
  10.1002/adfm.201602869} {\bibfield  {journal} {\bibinfo  {journal} {Advanced
  Functional Materials}\ }\textbf {\bibinfo {volume} {26}},\ \bibinfo {pages}
  {8643} (\bibinfo {year} {2016})}\BibitemShut {NoStop}%
\bibitem [{\citenamefont {Chang}\ \emph
  {et~al.}(2017{\natexlab{a}})\citenamefont {Chang}, \citenamefont {Avci},
  \citenamefont {Nikonov},\ and\ \citenamefont {Young}}]{Chang2017a}%
  \BibitemOpen
  \bibfield  {author} {\bibinfo {author} {\bibfnamefont {S.-C.}\ \bibnamefont
  {Chang}}, \bibinfo {author} {\bibfnamefont {U.~E.}\ \bibnamefont {Avci}},
  \bibinfo {author} {\bibfnamefont {D.~E.}\ \bibnamefont {Nikonov}}, \ and\
  \bibinfo {author} {\bibfnamefont {I.~A.}\ \bibnamefont {Young}},\ }\href
  {\doibase 10.1109/jxcdc.2017.2750108} {\bibfield  {journal} {\bibinfo
  {journal} {{IEEE} Journal on Exploratory Solid-State Computational Devices
  and Circuits}\ }\textbf {\bibinfo {volume} {3}},\ \bibinfo {pages} {56}
  (\bibinfo {year} {2017}{\natexlab{a}})}\BibitemShut {NoStop}%
\bibitem [{\citenamefont {B{\"o}scke}\ \emph {et~al.}(2011)\citenamefont
  {B{\"o}scke}, \citenamefont {M{\"u}ller}, \citenamefont {Br{\"a}uhaus},
  \citenamefont {Schr{\"o}der},\ and\ \citenamefont
  {B{\"o}ttger}}]{Boescke2011}%
  \BibitemOpen
  \bibfield  {author} {\bibinfo {author} {\bibfnamefont {T.~S.}\ \bibnamefont
  {B{\"o}scke}}, \bibinfo {author} {\bibfnamefont {J.}~\bibnamefont
  {M{\"u}ller}}, \bibinfo {author} {\bibfnamefont {D.}~\bibnamefont
  {Br{\"a}uhaus}}, \bibinfo {author} {\bibfnamefont {U.}~\bibnamefont
  {Schr{\"o}der}}, \ and\ \bibinfo {author} {\bibfnamefont {U.}~\bibnamefont
  {B{\"o}ttger}},\ }\href {\doibase 10.1063/1.3634052} {\bibfield  {journal}
  {\bibinfo  {journal} {Applied Physics Letters}\ }\textbf {\bibinfo {volume}
  {99}},\ \bibinfo {pages} {102903} (\bibinfo {year} {2011})}\BibitemShut
  {NoStop}%
\bibitem [{\citenamefont {M{\"u}ller}\ \emph {et~al.}(2012)\citenamefont
  {M{\"u}ller}, \citenamefont {B{\"o}scke}, \citenamefont {Schr{\"o}der},
  \citenamefont {Mueller}, \citenamefont {Br{\"a}uhaus}, \citenamefont
  {B{\"o}ttger}, \citenamefont {Frey},\ and\ \citenamefont
  {Mikolajick}}]{Mueller2012}%
  \BibitemOpen
  \bibfield  {author} {\bibinfo {author} {\bibfnamefont {J.}~\bibnamefont
  {M{\"u}ller}}, \bibinfo {author} {\bibfnamefont {T.~S.}\ \bibnamefont
  {B{\"o}scke}}, \bibinfo {author} {\bibfnamefont {U.}~\bibnamefont
  {Schr{\"o}der}}, \bibinfo {author} {\bibfnamefont {S.}~\bibnamefont
  {Mueller}}, \bibinfo {author} {\bibfnamefont {D.}~\bibnamefont
  {Br{\"a}uhaus}}, \bibinfo {author} {\bibfnamefont {U.}~\bibnamefont
  {B{\"o}ttger}}, \bibinfo {author} {\bibfnamefont {L.}~\bibnamefont {Frey}}, \
  and\ \bibinfo {author} {\bibfnamefont {T.}~\bibnamefont {Mikolajick}},\
  }\href {\doibase 10.1021/nl302049k} {\bibfield  {journal} {\bibinfo
  {journal} {Nano Letters}\ }\textbf {\bibinfo {volume} {12}},\ \bibinfo
  {pages} {4318} (\bibinfo {year} {2012})}\BibitemShut {NoStop}%
\bibitem [{\citenamefont {Sharma}\ \emph {et~al.}(2017)\citenamefont {Sharma},
  \citenamefont {Tapily}, \citenamefont {Saha}, \citenamefont {Zhang},
  \citenamefont {Shaughnessy}, \citenamefont {Aziz}, \citenamefont {Snider},
  \citenamefont {Gupta}, \citenamefont {Clark},\ and\ \citenamefont
  {Datta}}]{Sharma2017}%
  \BibitemOpen
  \bibfield  {author} {\bibinfo {author} {\bibfnamefont {P.}~\bibnamefont
  {Sharma}}, \bibinfo {author} {\bibfnamefont {K.}~\bibnamefont {Tapily}},
  \bibinfo {author} {\bibfnamefont {A.~K.}\ \bibnamefont {Saha}}, \bibinfo
  {author} {\bibfnamefont {J.}~\bibnamefont {Zhang}}, \bibinfo {author}
  {\bibfnamefont {A.}~\bibnamefont {Shaughnessy}}, \bibinfo {author}
  {\bibfnamefont {A.}~\bibnamefont {Aziz}}, \bibinfo {author} {\bibfnamefont
  {G.~L.}\ \bibnamefont {Snider}}, \bibinfo {author} {\bibfnamefont
  {S.}~\bibnamefont {Gupta}}, \bibinfo {author} {\bibfnamefont {R.~D.}\
  \bibnamefont {Clark}}, \ and\ \bibinfo {author} {\bibfnamefont
  {S.}~\bibnamefont {Datta}},\ }in\ \href {\doibase
  10.23919/vlsit.2017.7998160} {\emph {\bibinfo {booktitle} {2017 Symposium on
  {VLSI} Technology}}}\ (\bibinfo  {publisher} {{IEEE}},\ \bibinfo {year}
  {2017})\BibitemShut {NoStop}%
\bibitem [{\citenamefont {Kobayashi}\ \emph {et~al.}(2016)\citenamefont
  {Kobayashi}, \citenamefont {Ueyama}, \citenamefont {Jang},\ and\
  \citenamefont {Hiramoto}}]{Kobayashi2016}%
  \BibitemOpen
  \bibfield  {author} {\bibinfo {author} {\bibfnamefont {M.}~\bibnamefont
  {Kobayashi}}, \bibinfo {author} {\bibfnamefont {N.}~\bibnamefont {Ueyama}},
  \bibinfo {author} {\bibfnamefont {K.}~\bibnamefont {Jang}}, \ and\ \bibinfo
  {author} {\bibfnamefont {T.}~\bibnamefont {Hiramoto}},\ }in\ \href {\doibase
  10.1109/iedm.2016.7838402} {\emph {\bibinfo {booktitle} {2016 {IEEE}
  International Electron Devices Meeting ({IEDM})}}}\ (\bibinfo  {publisher}
  {{IEEE}},\ \bibinfo {year} {2016})\BibitemShut {NoStop}%
\bibitem [{\citenamefont {Chang}\ \emph {et~al.}(2018)\citenamefont {Chang},
  \citenamefont {Avci}, \citenamefont {Nikonov}, \citenamefont {Manipatruni},\
  and\ \citenamefont {Young}}]{Chang2018}%
  \BibitemOpen
  \bibfield  {author} {\bibinfo {author} {\bibfnamefont {S.-C.}\ \bibnamefont
  {Chang}}, \bibinfo {author} {\bibfnamefont {U.~E.}\ \bibnamefont {Avci}},
  \bibinfo {author} {\bibfnamefont {D.~E.}\ \bibnamefont {Nikonov}}, \bibinfo
  {author} {\bibfnamefont {S.}~\bibnamefont {Manipatruni}}, \ and\ \bibinfo
  {author} {\bibfnamefont {I.~A.}\ \bibnamefont {Young}},\ }\href {\doibase
  10.1103/physrevapplied.9.014010} {\bibfield  {journal} {\bibinfo  {journal}
  {Physical Review Applied}\ }\textbf {\bibinfo {volume} {9}} (\bibinfo {year}
  {2018}),\ 10.1103/physrevapplied.9.014010}\BibitemShut {NoStop}%
\bibitem [{\citenamefont {Yuan}\ \emph {et~al.}(2016)\citenamefont {Yuan},
  \citenamefont {Rizwan}, \citenamefont {Wong}, \citenamefont {Holland},
  \citenamefont {Anderson}, \citenamefont {Hook}, \citenamefont {Kienle},
  \citenamefont {Gadelrab}, \citenamefont {Gudem},\ and\ \citenamefont
  {Vaidyanathan}}]{Yuan2016}%
  \BibitemOpen
  \bibfield  {author} {\bibinfo {author} {\bibfnamefont {Z.~C.}\ \bibnamefont
  {Yuan}}, \bibinfo {author} {\bibfnamefont {S.}~\bibnamefont {Rizwan}},
  \bibinfo {author} {\bibfnamefont {M.}~\bibnamefont {Wong}}, \bibinfo {author}
  {\bibfnamefont {K.}~\bibnamefont {Holland}}, \bibinfo {author} {\bibfnamefont
  {S.}~\bibnamefont {Anderson}}, \bibinfo {author} {\bibfnamefont {T.~B.}\
  \bibnamefont {Hook}}, \bibinfo {author} {\bibfnamefont {D.}~\bibnamefont
  {Kienle}}, \bibinfo {author} {\bibfnamefont {S.}~\bibnamefont {Gadelrab}},
  \bibinfo {author} {\bibfnamefont {P.~S.}\ \bibnamefont {Gudem}}, \ and\
  \bibinfo {author} {\bibfnamefont {M.}~\bibnamefont {Vaidyanathan}},\ }\href
  {\doibase 10.1109/ted.2016.2602209} {\bibfield  {journal} {\bibinfo
  {journal} {{IEEE} Transactions on Electron Devices}\ }\textbf {\bibinfo
  {volume} {63}},\ \bibinfo {pages} {4046} (\bibinfo {year}
  {2016})}\BibitemShut {NoStop}%
\bibitem [{\citenamefont {Hoffmann}\ \emph {et~al.}(2018)\citenamefont
  {Hoffmann}, \citenamefont {Khan}, \citenamefont {Serrao}, \citenamefont {Lu},
  \citenamefont {Salahuddin}, \citenamefont {Pe{\v{s}}i{\'{c}}}, \citenamefont
  {Slesazeck}, \citenamefont {Schroeder},\ and\ \citenamefont
  {Mikolajick}}]{Hoffmann2018}%
  \BibitemOpen
  \bibfield  {author} {\bibinfo {author} {\bibfnamefont {M.}~\bibnamefont
  {Hoffmann}}, \bibinfo {author} {\bibfnamefont {A.~I.}\ \bibnamefont {Khan}},
  \bibinfo {author} {\bibfnamefont {C.}~\bibnamefont {Serrao}}, \bibinfo
  {author} {\bibfnamefont {Z.}~\bibnamefont {Lu}}, \bibinfo {author}
  {\bibfnamefont {S.}~\bibnamefont {Salahuddin}}, \bibinfo {author}
  {\bibfnamefont {M.}~\bibnamefont {Pe{\v{s}}i{\'{c}}}}, \bibinfo {author}
  {\bibfnamefont {S.}~\bibnamefont {Slesazeck}}, \bibinfo {author}
  {\bibfnamefont {U.}~\bibnamefont {Schroeder}}, \ and\ \bibinfo {author}
  {\bibfnamefont {T.}~\bibnamefont {Mikolajick}},\ }\href {\doibase
  10.1063/1.5030072} {\bibfield  {journal} {\bibinfo  {journal} {Journal of
  Applied Physics}\ }\textbf {\bibinfo {volume} {123}},\ \bibinfo {pages}
  {184101} (\bibinfo {year} {2018})}\BibitemShut {NoStop}%
\bibitem [{\citenamefont {Saha}\ \emph {et~al.}(2019)\citenamefont {Saha},
  \citenamefont {Ni}, \citenamefont {Dutta}, \citenamefont {Datta},\ and\
  \citenamefont {Gupta}}]{Saha2019}%
  \BibitemOpen
  \bibfield  {author} {\bibinfo {author} {\bibfnamefont {A.~K.}\ \bibnamefont
  {Saha}}, \bibinfo {author} {\bibfnamefont {K.}~\bibnamefont {Ni}}, \bibinfo
  {author} {\bibfnamefont {S.}~\bibnamefont {Dutta}}, \bibinfo {author}
  {\bibfnamefont {S.}~\bibnamefont {Datta}}, \ and\ \bibinfo {author}
  {\bibfnamefont {S.}~\bibnamefont {Gupta}},\ }\href {\doibase
  10.1063/1.5092707} {\bibfield  {journal} {\bibinfo  {journal} {Applied
  Physics Letters}\ }\textbf {\bibinfo {volume} {114}},\ \bibinfo {pages}
  {202903} (\bibinfo {year} {2019})}\BibitemShut {NoStop}%
\bibitem [{\citenamefont {Ishibashi}\ and\ \citenamefont
  {Takagi}(1971)}]{Ishibashi1971}%
  \BibitemOpen
  \bibfield  {author} {\bibinfo {author} {\bibfnamefont {Y.}~\bibnamefont
  {Ishibashi}}\ and\ \bibinfo {author} {\bibfnamefont {Y.}~\bibnamefont
  {Takagi}},\ }\href {\doibase 10.1143/jpsj.31.506} {\bibfield  {journal}
  {\bibinfo  {journal} {Journal of the Physical Society of Japan}\ }\textbf
  {\bibinfo {volume} {31}},\ \bibinfo {pages} {506} (\bibinfo {year}
  {1971})}\BibitemShut {NoStop}%
\bibitem [{\citenamefont {Kim}\ \emph {et~al.}(2017)\citenamefont {Kim},
  \citenamefont {Park}, \citenamefont {Hyun}, \citenamefont {Kim},
  \citenamefont {Kim}, \citenamefont {Lee}, \citenamefont {Moon}, \citenamefont
  {Lee}, \citenamefont {Park},\ and\ \citenamefont {Hwang}}]{Kim2017}%
  \BibitemOpen
  \bibfield  {author} {\bibinfo {author} {\bibfnamefont {Y.~J.}\ \bibnamefont
  {Kim}}, \bibinfo {author} {\bibfnamefont {H.~W.}\ \bibnamefont {Park}},
  \bibinfo {author} {\bibfnamefont {S.~D.}\ \bibnamefont {Hyun}}, \bibinfo
  {author} {\bibfnamefont {H.~J.}\ \bibnamefont {Kim}}, \bibinfo {author}
  {\bibfnamefont {K.~D.}\ \bibnamefont {Kim}}, \bibinfo {author} {\bibfnamefont
  {Y.~H.}\ \bibnamefont {Lee}}, \bibinfo {author} {\bibfnamefont
  {T.}~\bibnamefont {Moon}}, \bibinfo {author} {\bibfnamefont {Y.~B.}\
  \bibnamefont {Lee}}, \bibinfo {author} {\bibfnamefont {M.~H.}\ \bibnamefont
  {Park}}, \ and\ \bibinfo {author} {\bibfnamefont {C.~S.}\ \bibnamefont
  {Hwang}},\ }\href {\doibase 10.1021/acs.nanolett.7b04008} {\bibfield
  {journal} {\bibinfo  {journal} {Nano Letters}\ }\textbf {\bibinfo {volume}
  {17}},\ \bibinfo {pages} {7796} (\bibinfo {year} {2017})}\BibitemShut
  {NoStop}%
\bibitem [{\citenamefont {Aziz}\ \emph {et~al.}(2016)\citenamefont {Aziz},
  \citenamefont {Ghosh}, \citenamefont {Datta},\ and\ \citenamefont
  {Gupta}}]{Aziz2016}%
  \BibitemOpen
  \bibfield  {author} {\bibinfo {author} {\bibfnamefont {A.}~\bibnamefont
  {Aziz}}, \bibinfo {author} {\bibfnamefont {S.}~\bibnamefont {Ghosh}},
  \bibinfo {author} {\bibfnamefont {S.}~\bibnamefont {Datta}}, \ and\ \bibinfo
  {author} {\bibfnamefont {S.~K.}\ \bibnamefont {Gupta}},\ }\href@noop {}
  {\bibfield  {journal} {\bibinfo  {journal} {IEEE Electron Device Letters}\
  }\textbf {\bibinfo {volume} {37}},\ \bibinfo {pages} {805} (\bibinfo {year}
  {2016})},\ \bibinfo {note} {doi: \url{10.1109/LED.2016.2558149}}\BibitemShut
  {NoStop}%
\bibitem [{\citenamefont {Asai}\ \emph {et~al.}(2017)\citenamefont {Asai},
  \citenamefont {Fukuda}, \citenamefont {Hattori}, \citenamefont {Koike},
  \citenamefont {Miyata}, \citenamefont {Takahashi},\ and\ \citenamefont
  {Sakai}}]{Asai2017}%
  \BibitemOpen
  \bibfield  {author} {\bibinfo {author} {\bibfnamefont {H.}~\bibnamefont
  {Asai}}, \bibinfo {author} {\bibfnamefont {K.}~\bibnamefont {Fukuda}},
  \bibinfo {author} {\bibfnamefont {J.}~\bibnamefont {Hattori}}, \bibinfo
  {author} {\bibfnamefont {H.}~\bibnamefont {Koike}}, \bibinfo {author}
  {\bibfnamefont {N.}~\bibnamefont {Miyata}}, \bibinfo {author} {\bibfnamefont
  {M.}~\bibnamefont {Takahashi}}, \ and\ \bibinfo {author} {\bibfnamefont
  {S.}~\bibnamefont {Sakai}},\ }\href {\doibase 10.7567/jjap.56.04ce07}
  {\bibfield  {journal} {\bibinfo  {journal} {Japanese Journal of Applied
  Physics}\ }\textbf {\bibinfo {volume} {56}},\ \bibinfo {pages} {04CE07}
  (\bibinfo {year} {2017})}\BibitemShut {NoStop}%
\bibitem [{\citenamefont {Landau}(1937)}]{Landau1937}%
  \BibitemOpen
  \bibfield  {author} {\bibinfo {author} {\bibfnamefont {L.~D.}\ \bibnamefont
  {Landau}},\ }\href@noop {} {\bibfield  {journal} {\bibinfo  {journal} {Zh.
  Eksp. Teor. Fiz}\ }\textbf {\bibinfo {volume} {7}},\ \bibinfo {pages} {19}
  (\bibinfo {year} {1937})}\BibitemShut {NoStop}%
\bibitem [{\citenamefont {Ginzburg}(1945)}]{Ginzburg1945}%
  \BibitemOpen
  \bibfield  {author} {\bibinfo {author} {\bibfnamefont {V.~L.}\ \bibnamefont
  {Ginzburg}},\ }\href@noop {} {\bibfield  {journal} {\bibinfo  {journal} {Zh.
  Eksp. Teor. Fiz}\ }\textbf {\bibinfo {volume} {15}},\ \bibinfo {pages} {739}
  (\bibinfo {year} {1945})}\BibitemShut {NoStop}%
\bibitem [{\citenamefont {Devonshire}(1949)}]{Devonshire1949}%
  \BibitemOpen
  \bibfield  {author} {\bibinfo {author} {\bibfnamefont {A.}~\bibnamefont
  {Devonshire}},\ }\href@noop {} {\bibfield  {journal} {\bibinfo  {journal}
  {Phil. Mag}\ }\textbf {\bibinfo {volume} {40}},\ \bibinfo {pages} {1040}
  (\bibinfo {year} {1949})}\BibitemShut {NoStop}%
\bibitem [{\citenamefont {Hong}\ \emph {et~al.}(2008)\citenamefont {Hong},
  \citenamefont {Soh}, \citenamefont {Song},\ and\ \citenamefont
  {Lim}}]{Hong2008}%
  \BibitemOpen
  \bibfield  {author} {\bibinfo {author} {\bibfnamefont {L.}~\bibnamefont
  {Hong}}, \bibinfo {author} {\bibfnamefont {A.}~\bibnamefont {Soh}}, \bibinfo
  {author} {\bibfnamefont {Y.}~\bibnamefont {Song}}, \ and\ \bibinfo {author}
  {\bibfnamefont {L.}~\bibnamefont {Lim}},\ }\href {\doibase
  10.1016/j.actamat.2008.02.034} {\bibfield  {journal} {\bibinfo  {journal}
  {Acta Materialia}\ }\textbf {\bibinfo {volume} {56}},\ \bibinfo {pages}
  {2966} (\bibinfo {year} {2008})}\BibitemShut {NoStop}%
\bibitem [{\citenamefont {Li}\ \emph {et~al.}(2002)\citenamefont {Li},
  \citenamefont {Hu}, \citenamefont {Liu},\ and\ \citenamefont
  {Chen}}]{Li2002a}%
  \BibitemOpen
  \bibfield  {author} {\bibinfo {author} {\bibfnamefont {Y.~L.}\ \bibnamefont
  {Li}}, \bibinfo {author} {\bibfnamefont {S.~Y.}\ \bibnamefont {Hu}}, \bibinfo
  {author} {\bibfnamefont {Z.~K.}\ \bibnamefont {Liu}}, \ and\ \bibinfo
  {author} {\bibfnamefont {L.~Q.}\ \bibnamefont {Chen}},\ }\href {\doibase
  10.1063/1.1492025} {\bibfield  {journal} {\bibinfo  {journal} {Applied
  Physics Letters}\ }\textbf {\bibinfo {volume} {81}},\ \bibinfo {pages} {427}
  (\bibinfo {year} {2002})}\BibitemShut {NoStop}%
\bibitem [{\citenamefont {Nambu}\ and\ \citenamefont
  {Sagala}(1994)}]{Nambu1994}%
  \BibitemOpen
  \bibfield  {author} {\bibinfo {author} {\bibfnamefont {S.}~\bibnamefont
  {Nambu}}\ and\ \bibinfo {author} {\bibfnamefont {D.~A.}\ \bibnamefont
  {Sagala}},\ }\href {\doibase 10.1103/physrevb.50.5838} {\bibfield  {journal}
  {\bibinfo  {journal} {Physical Review B}\ }\textbf {\bibinfo {volume} {50}},\
  \bibinfo {pages} {5838} (\bibinfo {year} {1994})}\BibitemShut {NoStop}%
\bibitem [{\citenamefont {Tagantsev}(2008)}]{Tagantsev2008}%
  \BibitemOpen
  \bibfield  {author} {\bibinfo {author} {\bibfnamefont {A.~K.}\ \bibnamefont
  {Tagantsev}},\ }\href {\doibase 10.1080/00150190802437746} {\bibfield
  {journal} {\bibinfo  {journal} {Ferroelectrics}\ }\textbf {\bibinfo {volume}
  {375}},\ \bibinfo {pages} {19} (\bibinfo {year} {2008})}\BibitemShut
  {NoStop}%
\bibitem [{\citenamefont {Agarwal}\ \emph {et~al.}(2019)\citenamefont
  {Agarwal}, \citenamefont {Kushwaha}, \citenamefont {Lin}, \citenamefont
  {Kao}, \citenamefont {Liao}, \citenamefont {Dasgupta}, \citenamefont
  {Salahuddin},\ and\ \citenamefont {Hu}}]{Agarwal2019}%
  \BibitemOpen
  \bibfield  {author} {\bibinfo {author} {\bibfnamefont {H.}~\bibnamefont
  {Agarwal}}, \bibinfo {author} {\bibfnamefont {P.}~\bibnamefont {Kushwaha}},
  \bibinfo {author} {\bibfnamefont {Y.-K.}\ \bibnamefont {Lin}}, \bibinfo
  {author} {\bibfnamefont {M.-Y.}\ \bibnamefont {Kao}}, \bibinfo {author}
  {\bibfnamefont {Y.-H.}\ \bibnamefont {Liao}}, \bibinfo {author}
  {\bibfnamefont {A.}~\bibnamefont {Dasgupta}}, \bibinfo {author}
  {\bibfnamefont {S.}~\bibnamefont {Salahuddin}}, \ and\ \bibinfo {author}
  {\bibfnamefont {C.}~\bibnamefont {Hu}},\ }\href {\doibase
  10.1109/led.2019.2891540} {\bibfield  {journal} {\bibinfo  {journal} {{IEEE}
  Electron Device Letters}\ }\textbf {\bibinfo {volume} {40}},\ \bibinfo
  {pages} {463} (\bibinfo {year} {2019})}\BibitemShut {NoStop}%
\bibitem [{\citenamefont {Wang}\ and\ \citenamefont {Zhang}(2006)}]{Wang2006}%
  \BibitemOpen
  \bibfield  {author} {\bibinfo {author} {\bibfnamefont {J.}~\bibnamefont
  {Wang}}\ and\ \bibinfo {author} {\bibfnamefont {T.-Y.}\ \bibnamefont
  {Zhang}},\ }\href {\doibase 10.1103/physrevb.73.144107} {\bibfield  {journal}
  {\bibinfo  {journal} {Physical Review B}\ }\textbf {\bibinfo {volume} {73}}
  (\bibinfo {year} {2006}),\ 10.1103/physrevb.73.144107}\BibitemShut {NoStop}%
\bibitem [{\citenamefont {Dutta}\ \emph {et~al.}(2014)\citenamefont {Dutta},
  \citenamefont {Nikonov}, \citenamefont {Manipatruni}, \citenamefont {Young},\
  and\ \citenamefont {Naeemi}}]{Dutta2014}%
  \BibitemOpen
  \bibfield  {author} {\bibinfo {author} {\bibfnamefont {S.}~\bibnamefont
  {Dutta}}, \bibinfo {author} {\bibfnamefont {D.~E.}\ \bibnamefont {Nikonov}},
  \bibinfo {author} {\bibfnamefont {S.}~\bibnamefont {Manipatruni}}, \bibinfo
  {author} {\bibfnamefont {I.~A.}\ \bibnamefont {Young}}, \ and\ \bibinfo
  {author} {\bibfnamefont {A.}~\bibnamefont {Naeemi}},\ }\href {\doibase
  10.1109/tmag.2014.2320942} {\bibfield  {journal} {\bibinfo  {journal} {{IEEE}
  Transactions on Magnetics}\ }\textbf {\bibinfo {volume} {50}},\ \bibinfo
  {pages} {1} (\bibinfo {year} {2014})}\BibitemShut {NoStop}%
\bibitem [{\citenamefont {Chang}\ \emph
  {et~al.}(2017{\natexlab{b}})\citenamefont {Chang}, \citenamefont {Naeemi},
  \citenamefont {Nikonov},\ and\ \citenamefont {Gruverman}}]{Chang2017}%
  \BibitemOpen
  \bibfield  {author} {\bibinfo {author} {\bibfnamefont {S.-C.}\ \bibnamefont
  {Chang}}, \bibinfo {author} {\bibfnamefont {A.}~\bibnamefont {Naeemi}},
  \bibinfo {author} {\bibfnamefont {D.~E.}\ \bibnamefont {Nikonov}}, \ and\
  \bibinfo {author} {\bibfnamefont {A.}~\bibnamefont {Gruverman}},\ }\href
  {\doibase 10.1103/physrevapplied.7.024005} {\bibfield  {journal} {\bibinfo
  {journal} {Physical Review Applied}\ }\textbf {\bibinfo {volume} {7}}
  (\bibinfo {year} {2017}{\natexlab{b}}),\
  10.1103/physrevapplied.7.024005}\BibitemShut {NoStop}%
\bibitem [{\citenamefont {Chen}\ and\ \citenamefont {Shen}(1998)}]{Chen1998}%
  \BibitemOpen
  \bibfield  {author} {\bibinfo {author} {\bibfnamefont {L.}~\bibnamefont
  {Chen}}\ and\ \bibinfo {author} {\bibfnamefont {J.}~\bibnamefont {Shen}},\
  }\href {\doibase 10.1016/s0010-4655(97)00115-x} {\bibfield  {journal}
  {\bibinfo  {journal} {Computer Physics Communications}\ }\textbf {\bibinfo
  {volume} {108}},\ \bibinfo {pages} {147} (\bibinfo {year}
  {1998})}\BibitemShut {NoStop}%
\bibitem [{\citenamefont {Li}\ \emph {et~al.}(2017)\citenamefont {Li},
  \citenamefont {Huang}, \citenamefont {Hu},\ and\ \citenamefont
  {Zhang}}]{Li2017}%
  \BibitemOpen
  \bibfield  {author} {\bibinfo {author} {\bibfnamefont {G.}~\bibnamefont
  {Li}}, \bibinfo {author} {\bibfnamefont {X.}~\bibnamefont {Huang}}, \bibinfo
  {author} {\bibfnamefont {J.}~\bibnamefont {Hu}}, \ and\ \bibinfo {author}
  {\bibfnamefont {W.}~\bibnamefont {Zhang}},\ }\href {\doibase
  10.1103/physrevb.95.144111} {\bibfield  {journal} {\bibinfo  {journal}
  {Physical Review B}\ }\textbf {\bibinfo {volume} {95}} (\bibinfo {year}
  {2017}),\ 10.1103/physrevb.95.144111}\BibitemShut {NoStop}%
\bibitem [{sup()}]{supplemental_material}%
  \BibitemOpen
  \href@noop {} {\emph {\bibinfo {title} {\textup{See Supplemental Material for
  detailed simulations with spatially distributed bulk Landau
  coefficients.}}}}\BibitemShut {Stop}%
\end{thebibliography}%

\end{document}